\documentclass[%
 preprint, nolinenumbers,
 amsmath,amssymb,
 aps, physrev,
]{revtex4-2}
\usepackage{amssymb}
\usepackage{amsmath}
\usepackage{ytableau}
\usepackage{multirow}
\usepackage{caption}
\usepackage[margin=1in]{geometry}
\usepackage{booktabs}
\usepackage{graphicx}
\usepackage{dcolumn}
\usepackage{bm}

\begin{document}

\preprint{APS/123-QED}

\title{\textbf{Fisher information and quantum entropies of a 2D system under a non-central scalar and a vector potentials.} 
}%

\author{Ahmed Becir}
 
\author{Mustafa Moumni}%
 \email{Contact author: m.moumni@univ-batna.dz}
\affiliation{%
Laboratory of Radiations and Matter (LPRIM)\\ Physics Department, University of Batna I, Batna, Algeria }%
\altaffiliation[Also at ]{Laboratory of Photonic Physics and NanoMaterials, Department of Matter Sciences, University of Biskra, Algeria.}
\date{\today}

\begin{abstract}
We study a two-dimensional system influenced by a non-central potential consisting of a Kratzer potential with a dipole moment,  along with a vector potential of the Aharonov-Bohm (AB) effect.  We explore various information-theoretic measures, including Fisher information, Shannon entropy, Tsallis entropy, and Renyi entropy.  Our numerical results show that Fisher information increases with an increase in dissociation energy and decreases with increasing dipole moment, Aharonov-Bohm potential strength, and radial and angular quantum numbers. In contrast, the Shannon entropy, the Tsallis entropy, and the Renyi entropy decrease with rising dissociation energy, while they increase with an increase in dipole moment, Aharonov–Bohm potential strength, as well as the radial and angular quantum numbers. These observations collectively indicate that the precision and localization of particles in space are enhanced by increasing the dissociation energy and reduced when the dipole moment, Aharonov–Bohm potential strength, and both radial and angular quantum numbers are increased. 
\begin{description}
\item[Keywords]
Fisher information, Quantum entropies, non-central Kratzer potential, Aharonov-Bohm potential.
 
\end{description}
\end{abstract}

\maketitle


\section{\label{sec:1}Introduction:}

Quantum information is, as its name indicates, information on the state of a quantum system, and it combines principles from quantum mechanics and information theory to understand and manipulate information using quantum systems. Quantum information has some basic measures such as the Renyi and the Shannon entropies \cite{Cover2005,Uffink1990,Hall1999}, Fisher information \cite{Fisher1925,Frieden2004,Hall2000,Hall2001}, Tsallis entropy \cite{Portes2004,Grau2010,Santos1997}, Fisher-Shannon complexity \cite{Angulo2008,Montgomery2008,Manzano2012}, and Cramer–Rao complexity \cite{Angulo2008a}.

Quantum information is a rapidly evolving field with the potential to cause revolutionary advances in various scientific and technological domains. Several research studies have been conducted on the entropies and complexities of quantum systems. Position and momentum space entropies have been derived for various systems, including the isotropic harmonic oscillator and the hydrogen atom in D dimensions \cite{Yanez1994,Dehesa2006,Dehesa2005}, power-type potentials \cite{Dehesa2002}, Morse and Pöschl–Teller potentials \cite{Romera2006,Dehesa2006a}, as well as Dirac-delta-like quantum potentials \cite{Bouvrie2011}. Furthermore, uncertainty relations have been verified for modified isotropic harmonic oscillators and Coulomb potentials in \cite{Patil2007} and quantum information entropies and squeezing associated with the eigenstates of the isotonic oscillator have been discussed in \cite{Ghasemi2011}.

Recent researches had extended these investigations to encompass quantum information-entropic measures for half-line Coulomb potential \cite{Omiste2010}, ring-shaped modified Kratzer potential \cite{Yahia2014}, Pseudoharmonic Potential \cite{Yahia2015}, Eckart Manning Rosen Potential \cite{Onate2017}, Shifted Tietz-Wei Potential \cite{Onate2018}, Hulthen–Kratzer potential \cite{Obu2019}, Generalized Morse potential \cite{Onate2018a}, exponential-type potential \cite{Ikot2020}, screened Kratzer potential \cite{Amadi2020}, and for Eckart-Hellmann potential \cite{Inyang2022}.

In this study, our focus lies on examining some measures of quantum information for a two-dimensional system under the influence of a non-central potential consisting of a Kratzer potential and a dipole moment \cite{Heddar2019}, along with a vector potential of the Aharonov–Bohm (AB) effect. The Kratzer potential is significant when depicting the internuclear vibrations observed in diatomic molecules \cite{Ikhdair2011}. Its relevance extends to the fields of molecular spectroscopy and quantum chemistry \cite{VanHooydonk2008, Berkdemir2006}, as well as to the investigation of optical properties in semiconductor quantum dots and also 2D dichalcogenides materials \cite{Batra2018}.

The interest in two-dimensional (2D) materials has grown significantly because of their unique properties and diverse applications in various fields. Graphene has been extensively developed for many electronic applications due to its electronic, thermal, optical, and mechanical characteristics \cite{Sang2019}. It has shown potential in improving the performance of lithium-ion batteries by increasing their capacity \cite{Du2023} and has been used in the fabrication of wind and solar cells \cite{Bagade2023}. In addition, graphene appears promising as a hydrogen storage material \cite{Dwivedi2022}. Black phosphorus (BP), another outstanding 2D material, has attracted attention due to its role as a layered semiconductor with a tunable bandgap and high carrier mobility. It is considered one of the most promising candidates for numerous applications ranging from transistors to photonics, optoelectronics, sensors, batteries, and catalysis \cite{Xia2019, Zehan2021}. Moreover, transition-metal dichalcogenides, which are 2D materials, have semiconducting properties and have many applications in electronics, spintronics, optoelectronics, energy harvesting, flexible electronics, DNA sequencing, and personalized medicine \cite{Manzeli2017}.

This paper is organized as follows. In Section ~\ref{sec:2} we solve the Schrödinger equation for a 2D system under the influence of a non-central potential composed of a Kratzer potential along with a dipole moment, as well as a vector potential from the Aharonov–Bohm (AB) effect, and we find the exact energy eigenvalues and the corresponding normalized wave functions. Sections ~\ref{sec:3} and ~\ref{sec:4} are devoted to analytically deriving the Fisher information measure, Shannon, Tsallis, and Renyi entropies of the given system. In Section ~\ref{sec:5}, we explore the impact of the dissociation energy, the dipole moment, the AB field, and also the radial and the angular quantum numbers on the derived quantum information measures, and subsequently we discuss the obtained results. Finally, the last Section ~\ref{sec:6} provides a concise conclusion summarizing our findings.
\section{\label{sec:2}The Exact solution of 2D Schrödinger equation with non-central scalar and vector potentials}
The 2D Schrödinger equation for a system subjected to both a scalar potential and a vector potential of the Aharonov-Bohm (AB) effect, $\vec{\phi}_{AB}$ is written as \cite{Baazouzi2020}:
\begin{equation}
\left[\frac{1}{2\mu} \left(i\hbar\vec{\nabla}+e\vec{\phi}_{AB}\right)^2+qV(r,\theta)\right]\psi(r,\theta) = E\psi(r,\theta) : \quad \vec{\phi}_{AB} = \frac{\phi_{AB}}{2\pi r}\vec{e}_{\theta}
\label{eq:1}.
\end{equation}
Since the AB field satisfies the Coulomb gauge $\vec{\nabla} \cdot \vec{\phi}_{AB}=0$, then:
\begin{equation}
\left(i\hbar \vec{\nabla} + e\vec{A}_{AB}\right)^2 \psi(r, \theta) = \left(-\hbar^2\Delta + e^2 A^2_{AB} + 2ie\hbar \vec{A}_{AB} \cdot \vec{\nabla}\right) \psi(r, \theta)
\label{eq:2}.
\end{equation}
and the Schrödinger Eq.~(\ref{eq:2}) becomes:
\begin{equation}
\left[-\frac{\hbar^2}{2\mu}\Delta+\frac{e^2\phi^2_{AB}}{8\pi^2\mu r^2}+i\frac{e\hbar\phi_{AB}}{2\mu\pi r^2}\frac{\partial}{\partial{\theta}}+V(r,\theta)\right]\psi(r,\theta)= E\psi(r,\theta)
\label{eq:3}.
\end{equation}
The non-central potential studied in the article consists of two components: a modified Kratzer potential and an angular dipole moment \cite{Heddar2019}. The former is more appropriate than the Coulomb potential for the study of 2D materials such as dichalcogenides \cite{Molas2019}, and the later is added to account for the possibility of a non-symmetric charge distribution:
\begin{equation}
V\left( {r,\theta } \right) = D_e\left(\frac{r-r_e}{r}\right)^2 + \frac{D \cos \theta}{r^2}=\frac{A}{r}+\frac{B}{r^2}+C+\frac{D \cos \theta}{r^2}
\label{eq:4}
\end{equation}
where $A=-2r_e D_e, B=r_e^2 D_e, C=D_e$, $D_e$ is the dissociation energy, which represents the energy required to completely separate the two atoms, $r_e$ is the equilibrium bond length, which is the distance at which the potential energy is minimized, and $D$ is the dipole moment.\newline
Substituting eq.(4) into eq.(3) yields:
\begin{align}
& \left[-\frac{\hbar^2}{2\mu}\Delta + \frac{e^2\phi^2_{AB}}{8\pi^2\mu r^2} + i\frac{e\hbar\phi_{AB}}{2\mu\pi r^2}\frac{\partial}{\partial{\theta}} \right. \nonumber \\
& \left. + \left(\frac{A}{r} + \frac{B}{r^2} + \frac{D\cos \theta}{r^2}\right)\right]\psi(r, \theta) = (E-C)\psi(r, \theta)
\label{eq.5}.
\end{align}
If we let $E_r=2\mu\hbar^{-2}(E-C)$ , $\gamma=B+\frac{\hbar^2}{2\mu}\frac{\phi^2_{AB}}{\phi^2_0}: \phi_0=\frac{h}{e}$, then we get
\begin{equation}
\left[\frac{\partial^2}{\partial r^2} + \frac{1}{r}\frac{\partial}{\partial r} - \frac{2\mu}{\hbar^2}\frac{A}{r} - \frac{2\mu}{\hbar^2}\frac{\gamma}{r^2} + \frac{1}{r^2}\left(\frac{\partial^2}{\partial \theta^2} -
2i\frac{\phi_{AB}}{\phi_0}\frac{\partial}{\partial \theta} -\frac{2\mu}{\hbar^2} D\cos\theta\right)\right]\psi = -E_r\psi
\label{eq:6}.
\end{equation}
Now making use of $\psi (r,\theta ) = {r^{ - 1/2}}R(r)\Theta (\theta )$, Eq.~(\ref{eq:6}) can be decoupled into radial and angular parts:
\begin{equation}
\left(\frac{d^2}{d \theta^2} - 2i\frac{\phi_{AB}}{\phi_0}\frac{d}{d \theta} - \frac{2\mu}{\hbar^2}D\cos \theta \right) \Theta(\theta) = E_\theta \Theta(\theta)
\label{eq:7}
\end{equation}
\begin{equation}
\left(\frac{d^2}{d r^2} + \left(E_\theta + \frac{1}{4} - \frac{2\mu}{\hbar^2}\gamma\right)\frac{1}{r^2} - \frac{2\mu}{\hbar^2}\frac{A}{r} + E_r \right) R(r) = 0
\label{eq:8}
\end{equation}
Firstly, one needs to solve the angular Eq.~(\ref{eq:7}) to determine the angular eigenvalue $E_\theta$. Subsequently, this value should be substituted into the radial Eq.~(\ref{eq:8}) to deduce the energy eigenvalue \cite{Heddar2019, Baazouzi2020}. By expressing the angular solution as $\Theta(\theta)=e^{i\delta\theta}\Phi(\theta)$, where $\delta=\frac{\phi_{AB}}{\phi_0}$, and defining $\theta  = 2z$, $a = 4\left(\delta^2-E_\theta \right)$, and $b = \frac{4\mu}{\hbar^2} D$, the angular eq.(7) transforms into the Mathieu equation:
\begin{equation}
\frac{\partial^2 \Phi(z)}{\partial z^2} + (a - 2b\cos 2z) \Phi(z) = 0
\label{eq:9}
\end{equation}
The differential Eq.~(\ref{eq:9}) has periodic solutions of periods $\pi$ or $2\pi$, represented by the cosine-elliptic $c{e_{2m}}(z)$ and the sine-elliptic $s{e_{2m}}(z)$ functions where $m$ is a positive integer \cite{AbraSteg1972}. If we keep the parameter $b$ fixed, the Mathieu solutions are periodic only for specific values of the other parameter $a$. The latter parameter is called the characteristic number and is given for fractional $m$ by \cite{Heddar2019, Baazouzi2020, Jazar2020, Frenkel2001}
\begin{equation}
a_{2m} \approx 4m^2 + \frac{1}{2l}b^2 + \frac{20m^2 + 7}{32l^3(l - 3)}b^4 + \frac{36m^4 + 232m^2 + 29}{64l^5(l - 3)(l - 8)}b^6 + \ldots,
\quad \text{with } l = 4m^2 - 1
\label{eq:10}
\end{equation}
The phase factor $\exp(i\delta\theta)$ in $\Theta(\theta)$ arising from the AB field induces a shift in the angular quantum number $m$ to $m+\delta$ \cite{Heddar2019, Baazouzi2020, Gadella2011}. Consequently, the corresponding angular eigenvalue becomes:
\begin{equation}
E_\theta ^{m,\delta} = \delta^2 - \frac{1}{4}c_{2(m+\delta)}\left( \frac{4\mu}{\hbar^2} D \right)
\label{eq:11}
\end{equation}
Now, considering the asymptote limits for the radial Eq.~(\ref{eq:8}), the radial eigenfunction can be expressed as $R(r) = r^\lambda e^{-\beta r}f(r)$. Substituting this expression into Eq.~(\ref{eq:8}), we get:
\begin{equation}
\left( r\frac{d^2}{dr^2} + 2\left( \lambda - \beta r \right)\frac{d}{dr} - 2\left( \frac{\mu A}{\hbar^2} + \lambda \beta \right) \right) f(r) = 0
\label{eq:12}
\end{equation}
In deriving Eq.~(\ref{eq:12}), we eliminated the coefficients of the terms involving $r$ and $\frac{1}{r}$, knowing that $\beta$ and $\lambda$ are free parameters. This was achieved by setting:
\begin{equation}
 \beta=\sqrt{-E_r}=\sqrt{-\frac{2\mu}{\hbar^2}(E-C)}
 \label{eq:13}
\end{equation}
\begin{equation}
 \lambda=\frac{1}{2}+\sqrt{-E_\theta^{(m,\delta)}+\frac{2\mu}{\hbar^2}\gamma}=\frac{1}{2}+\sqrt{-E_\theta^{(m,\delta)}+\frac{2\mu}{\hbar^2}B+\frac{\phi_{AB}^2}{\phi_0^2}}
 \label{eq:14}
\end{equation}
By defining $x = 2\beta r$, the differential equation~(\ref{eq:12}) is transformed to a confluent hypergeometric equation:
\begin{equation}
\left( x\frac{d^2}{dx^2} + \left( 2 \lambda - x \right)\frac{d}{dx} - \left( \frac{\mu A}{\hbar^2\beta}+\lambda \right) \right) f(x) = 0
\label{eq.15}
\end{equation}
with solutions:
\begin{equation}
f(x) = \mathcal{N} {}_1F_1\left(-n_r, 2\lambda, x\right)
\label{eq:16}
\end{equation}
where $- n_r= \frac{\mu A}{\hbar^2\beta}+\lambda $ is the condition of quantization obtained from the asymptotic behavior of the confluent series ($r \rightarrow \infty \Rightarrow {}_1F_1 = 0$). \newline
The wavefunction of the system is then given by:
\begin{equation}
\psi (r,\theta) = \mathcal{N} e^{i\delta\theta}r^{\lambda - \frac{1}{2}} e^{-\beta r} \Phi(\theta) {}_1F_1\left( -n_r, 2\lambda, 2\beta r \right)
\label{eq:17}
\end{equation}
with energy eigenvalues:
\begin{equation}
E_{{n_r},m} =  -\frac{1}{2}\left[ \sqrt{\frac{\hbar^2}{\mu}}\frac{1}{A}\left( n_r + \frac{1}{2} + \sqrt{-E_\theta^{(m,\delta)}+\frac{2\mu}{\hbar^2}B+\frac{\phi_{AB}^2}{\phi_0^2}} \right)\right]^{-2}
\label{eq:18}
\end{equation}
The wavefunction~(\ref{eq:17}) can be rewritten in term of Laguerre polynomials as:
\begin{equation}
\psi (x,\theta ) = N x^{\lambda - \frac{1}{2}} e^{-\frac{x}{2}} e^{i\delta\theta}\Phi (\theta) L_{n_r}^{2\lambda - 1}\left(x\right), \text{ with } x = 2\beta r
\label{eq:19}
\end{equation}
with the normalization constant:
\begin{equation}
N = \sqrt{\frac{{2{\beta ^2}n!}}{{\left( {n + 2\lambda  - 1} \right)!(n + \lambda )\pi }}}
\label{eq:20}
\end{equation}
where we have used the fact that (\cite{Moll2014})
\begin{equation}
L_{n}^{\lambda-1}(x)\thicksim {}_1{F_1}\left( { - n,\lambda ,x} \right)
\label{eq:21}
\end{equation}
\section{\label{sec:3}Fisher Information and Shannon Entropy}
\subsection{\label{subsec:3a}Fisher Information}
Fisher information is a method for quantifying the extent of information that an observable random variable $X$ provides regarding an unknown parameter $\theta$, and it has many applications in statistics and information theory \cite{CaseBerg2001, Sakurai2020}. It primarily focuses on capturing local variations within the density function and its measure is formally defined as follows \cite{Kali2011}:
\begin{equation}
I(\rho) = \int \frac{\left(\vec{\nabla}\rho(\vec{r})\right)^2}{\rho(\vec{r})} \, d\vec{r}
\label{eq:22}
\end{equation}
In 2D dimension, taking the gradient operator in polar coordinates$\vec{\nabla} = \left( \frac{\partial }{\partial r},\frac{1}{r}\frac{\partial }{\partial \theta } \right)$, the expression for the Fisher information is as follows:
\begin{equation}
I\left( \rho  \right) = \int {\frac{1}{\rho \left( {\vec r} \right)} {\left[ \frac{\partial \rho \left( {\vec r} \right)}{\partial r} \right]^2}d\vec r} + \int {\frac{1}{\rho \left( {\vec r} \right)} {\left[ \frac{\partial \rho \left( {\vec r} \right)}{r\partial \theta } \right]^2}d\vec r} \equiv {I_1} + {I_2}
\label{eq:23}
\end{equation}
where $\rho \left( {\vec r} \right) \equiv \rho \left( {r,\theta } \right) = {\left| {\psi (r,\theta )} \right|^2}$ is the probability density.
Using the wave function~(\ref{eq:19}) along with the normalization constant~(\ref{eq:20}), the probability density is expressed as:
\begin{equation}
\rho(r, \theta) = N^2 x^{2\lambda - 1} e^{-x} \Phi^2(\theta) [L_n^{2\lambda - 1}(x)]^2, \text{ with } x = 2\beta r
\label{eq:24}
\end{equation}
The derivative in the Eq.~(\ref{eq:23}) yields:
\begin{align}
\frac{\partial \rho(r,\theta)}{\partial r} =& 2\beta N_r\Phi^2(\theta)\frac{\partial}{\partial x}\left(x^{2\lambda - 1}e^{-x}[L_n^{2\lambda - 1}(x)]^2\right)\nonumber \\
&= 2\beta N^2\Phi^2(\theta)x^{2\lambda - 1}e^{-x}\left[\left(\frac{2\lambda - 1}{x} - 1\right)[L_n^{2\lambda - 1}(x)]^2 - 2L_n^{2\lambda - 1}(x)L_{n - 1}^{2\lambda}(x)\right]
\label{eq:25}
\end{align}
and so that:
\begin{align}
I_1 &= \int\limits_0^\infty \int\limits_0^{2\pi} \frac{1}{\rho(r,\theta)}\left[\frac{\partial \rho(r,\theta)}{\partial r}\right]^2 r\,dr\,d\theta \nonumber\\
&= \pi N^2 \int\limits_0^\infty x^{2\lambda} e^{-x}\left[\left(\frac{2\lambda - 1}{x} - 1\right)L_n^{2\lambda - 1}(x) - 2L_{n - 1}^{2\lambda}(x)\right]^2 dx\nonumber\\
&=\frac{2\beta^2}{n+\lambda}\left[4n-(2\lambda-1)\right]+4\beta^2
\label{eq:26}
\end{align}	
where we have used the fact that$\int\limits_0^{2\pi} \Phi^2(\theta) \, d\theta = \pi$ and the relation: \cite{Nieto1979}
\begin{align}
&\int\limits_0^\infty x^{\alpha+\beta} e^{-x}\left[L_n^{\alpha}(x)\right]^2\, dx \nonumber\\
&= \frac{\Gamma(\alpha+n+1)}{\Gamma(n+1)} \sum_{k=0}^{n} (-1)^k \frac{\Gamma(n-k-\beta)}{\Gamma(-k-\beta)} \nonumber\\
&\quad\times \frac{\Gamma(\alpha+k+\beta+1)}{\Gamma(\alpha+k+1)} \frac{1}{\Gamma(k+1)\Gamma(n-k+1)} \quad \text{with } \operatorname{Re}(\alpha+\beta+1)>1
\label{eq.27}
\end{align}
The second integral in Eq.~(\ref{eq:23}) involves the following integral:
\begin{equation}
\int\limits_0^{2\pi} \frac{1}{\Phi^2(\theta)} \left[\frac{\partial \Phi^2(\theta)}{\partial \theta}\right]^2 \, d\theta = 4\int\limits_0^{2\pi} \left[\frac{\partial \Phi(\theta)}{\partial \theta}\right]^2 \, d\theta = 4m^2\pi
\label{eq:28}
\end{equation}
Computing the integral Eq.~(\ref{eq:31}) is straightforward if the characteristic number $b=0$. For $b\neq0$, it is also easy to check that this integral tends to $4m^2\pi$ for all values of $m$.\newline
Using the result Eq.~(\ref{eq:28}), the second integral in Eq.~(\ref{eq:23}) becomes:
\begin{align}
I_2&= 4\pi m^2 N^2 \int\limits_0^\infty x^{2\lambda - 2} e^{-x} [L_n^{2\lambda - 1}(x)]^2 \, dx \nonumber\\
&= \frac{8m^2\beta^2}{(n + \lambda)(2\lambda - 1)}
\label{eq:29}
\end{align}
Therefore, the Fisher information is given by:
\begin{equation}
I(\rho) = I_1+I_2=\frac{2\beta^2}{(n + \lambda)} \left[4n + \frac{4m^2}{2\lambda - 1} - (2\lambda - 1)\right] + 4\beta^2
\label{eq:30}
\end{equation}
\begin{figure}
  \centering
  \includegraphics[width=0.6\textwidth]{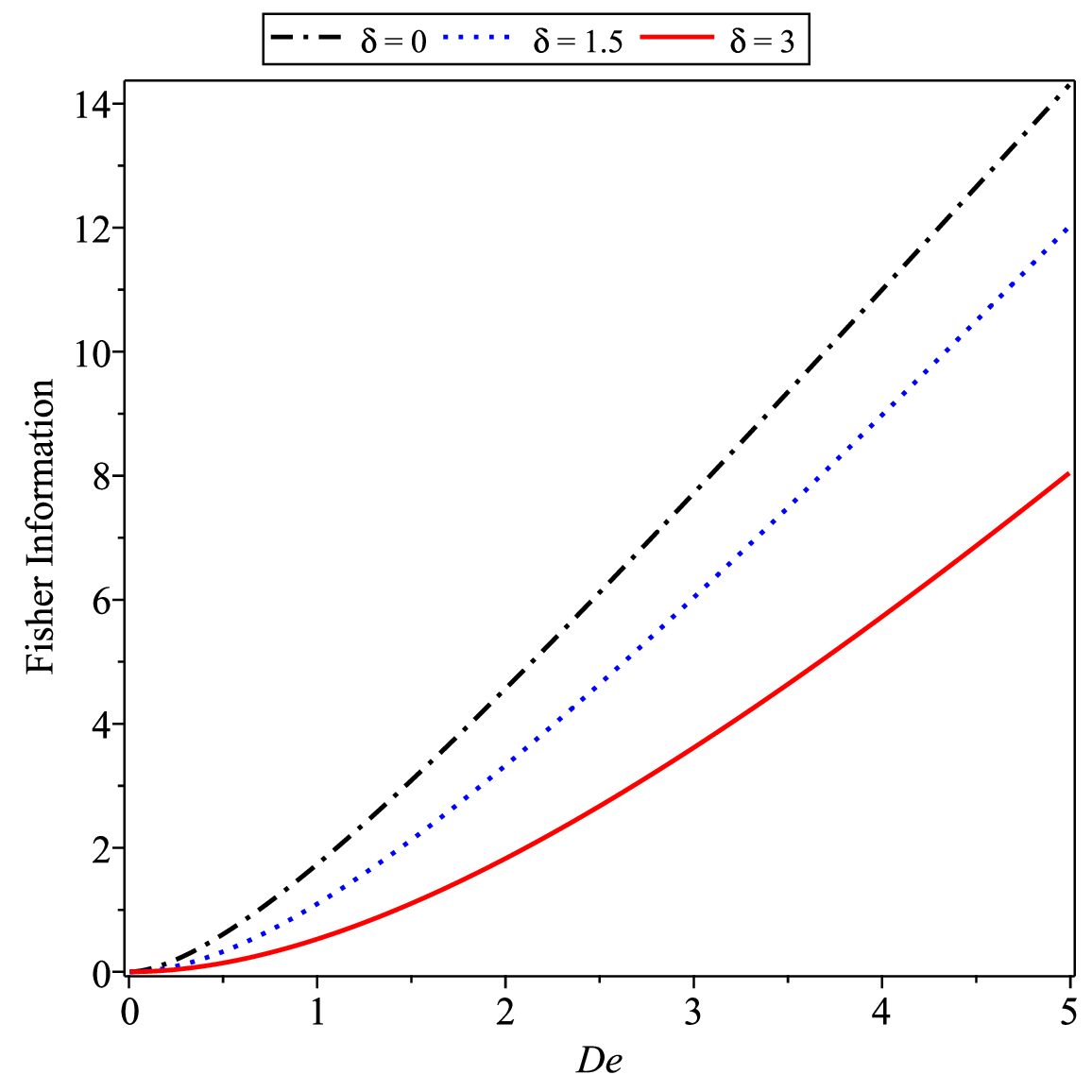}
  \caption{Fisher information versus the Dissociation energy for $D=0$, $r_e=1$ (All quantities are in atomic units), with $n=2$ and $m=0$.}
  \label{fig:1}
\end{figure}
\begin{figure}
  \centering
  \includegraphics[width=0.6\textwidth]{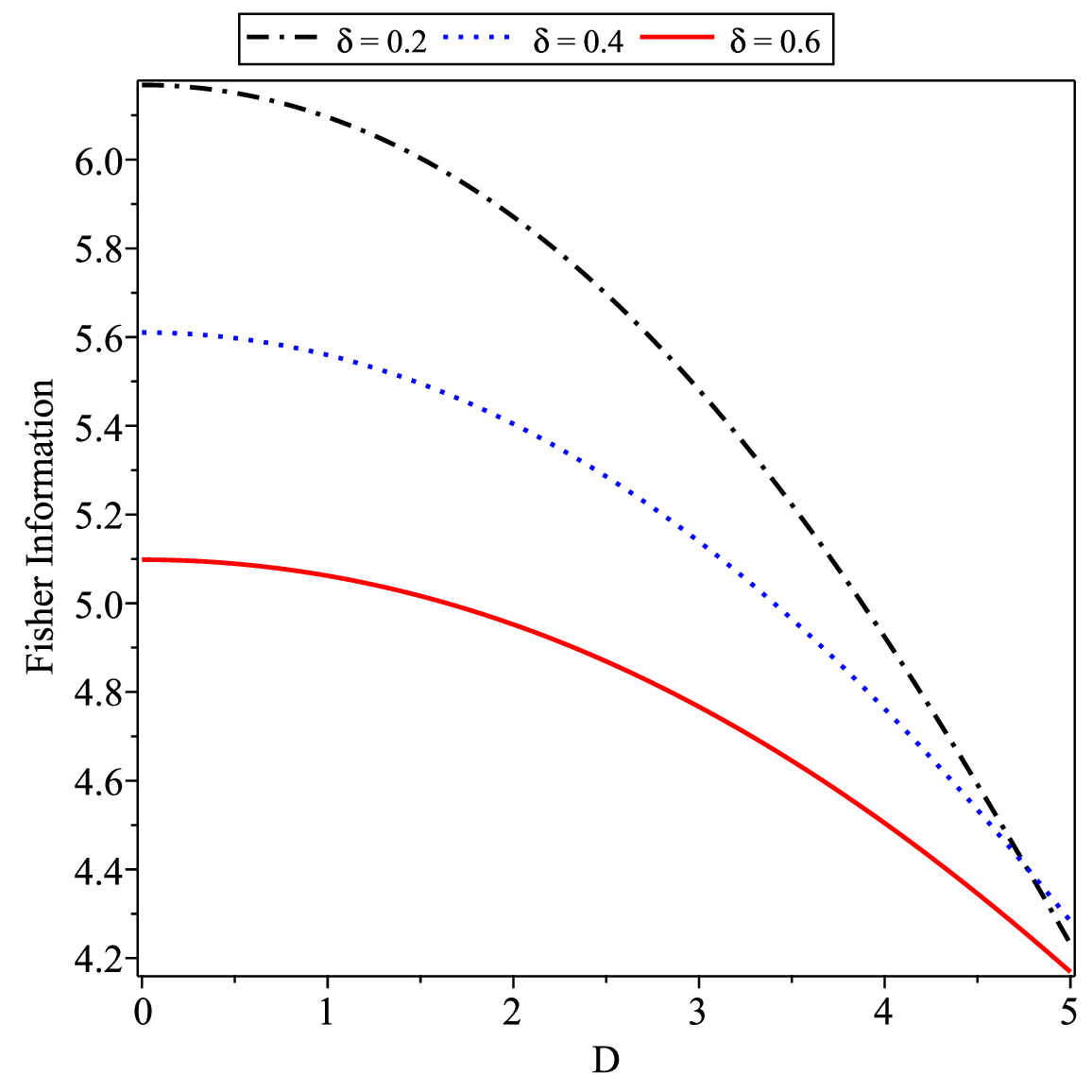}
  \caption{Fisher information versus the Dipole moment for $De=3$, $r_e=1$ (All quantities are in atomic units), with $n=2$ and $m=2$.}
  \label{fig:2}
\end{figure}
\subsection{\label{subsec:3b}Shannon Information}
The Shannon entropy quantifies the uncertainty associated with the localization of a particle in space. Lower entropy values correspond to greater precision in predicting the particle's location. The Shannon information entropy is defined as:\cite{Kali2011}
\begin{align}
S (\rho) &=-\int \rho(\vec{r}) \log \rho(\vec{r}) \, d\vec{r} \nonumber\\
&= -\frac{N^2}{4\beta^2} \int\limits_0^\infty \int\limits_0^{2\pi} \Phi(\theta)^2 x^{2\lambda} e^{-x} \left[L_n^{2\lambda - 1}(x)\right]^2 \log \left[N_r \Phi(\theta)^2 x^{2\lambda - 1} e^{-x} \left[L_n^{2\lambda - 1}(x)\right]^2\right] \, dx \, d\theta \nonumber\\
&= -\frac{\pi N^2}{4\beta^2} \int\limits_0^\infty x^{2\lambda} e^{-x} \left[L_n^{2\lambda - 1}(x)\right]^2 \log \left[N^2 x^{2\lambda - 1} e^{-x} \left[L_n^{2\lambda - 1}(x)\right]^2\right] \, dx \nonumber\\
&- \frac{I_\theta N^2}{4\beta^2} \int\limits_0^\infty x^{2\lambda} e^{-x} \left[L_n^{2\lambda - 1}(x)\right]^2 \, dx
\label{eq:31}
\end{align}
with $x=2\beta r$ and:
\begin{equation}
I_\theta = \int\limits_0^{2\pi} \Phi(\theta)^2 \log(\Phi(\theta)^2) \, d\theta=\pi\left(1-2\ln(2)\right)
\label{eq:32}
\end{equation}
where the logarithm is taken in base $e$ since we are interested in natural units. The integral~(\ref{eq:32}) is easily obtained for the case of $b=0$. For $b\neq0$ one can also show that the integral tends to the same value~(\ref{eq:32}). The integral Eq.~(\ref{eq:31}) can be decomposed into four separate integrals:
\begin{equation}
S_1 = -\frac{{I_\theta N^2}}{{4 \beta^2}} \int_0^\infty x^{2\lambda} e^{-x} \left[L_n^{2\lambda - 1}(x)\right]^2 \, dx = -\frac{I_\theta}{\pi}
\label{eq:33}
\end{equation}
For the other three integral parts, it is convenient to introduce the orthonormal Laguerre polynomials:
\begin{equation}
\tilde{L}_n^{\lambda}(x)=\left[\frac{\Gamma(n+1)}{\Gamma(\lambda+n+1)}\right]^{\frac{1}{2}}L_n^{\lambda}(x)
\label{eq:34}
\end{equation}
because they satisfy the simplified orthonormal relation:
\begin{equation}
 \int_{0}^{\infty} \tilde{L}_n^\lambda(x)\tilde{L}_m^\lambda(x) e^{-x} x^\lambda \, dx = \delta_{nm}
 \label{34}
\end{equation}
The remaining integrals are then given by:
\begin{equation}
S_2= -\frac{\pi N^2}{{4 \beta^2}} \int_0^\infty x^{2\lambda} e^{-x} \left[L_n^{2\lambda - 1}(x)\right]^2 \log\left(N^2\frac{\Gamma(n+2\lambda)}{\Gamma(n+1)}\right) \, dx = -\log\left(\frac{2\beta^{2}}{(n+\lambda)\pi}\right)
\label{eq:36}
\end{equation}
\begin{align}
S_3 &= -\frac{\pi N^2}{4 \beta^2} \int_0^\infty x^{2\lambda} e^{-x} \left[L_n^{2\lambda - 1}(x)\right]^2 \log\left(x^{2\lambda-1} e^{-x}\right) \, dx \nonumber\\
&= \frac{(n + 2\lambda) \left[-(2\lambda - 1)\Psi(n + 2\lambda + 1) + 2n + 2\lambda + 1\right]}{2 (n + \lambda)}
\label{eq:37}
\end{align}
where $\Psi(z)=\frac{\Gamma^{'}(z)}{\Gamma(z)}$ is the Digamma function \cite{Sanchez2000}.
\begin{align}
S_4 &= -\frac{\pi N^2}{{4 \beta^2}} \frac{\Gamma(n+2\lambda)}{\Gamma(n+1)}\int_0^\infty x^{2\lambda} e^{-x} \left[\tilde{L}_n^{2\lambda - 1}(x)\right]^2 \log \left(\left[\tilde{L}_n^{2\lambda - 1}(x)\right]^2\right) \, dx\nonumber\\
&=-\frac{1}{2(n+\lambda)}\left(-6n^2 + 4\lambda n\log n + 2n\left[\log(2\pi) - 4\lambda - 2\right] + O(n)\right)
\label{eq:38}
\end{align}
The last integral is derived using the $p$-norm method \cite{Dehesa1998} at the asymptotic limits.\newline
\begin{figure}
  \centering
  \includegraphics[width=0.6\textwidth]{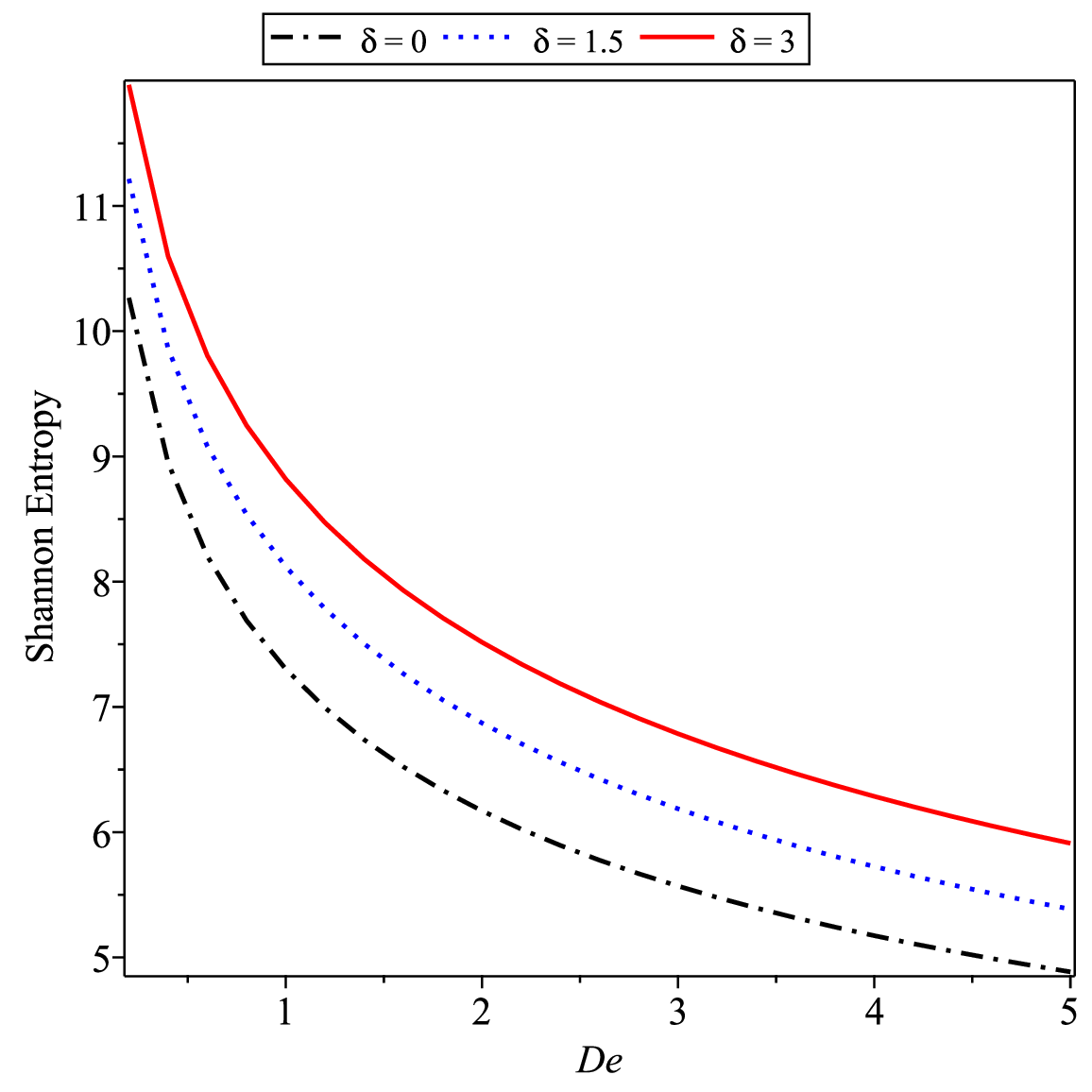}
  \caption{Shannon entropy versus the Dissociation energy for $D=0$, $r_e=1$ (All quantities are in atomic units), with $n=2$ and $m=0$.}
  \label{fig:3}
\end{figure}
\begin{figure}
  \centering
  \includegraphics[width=0.6\textwidth]{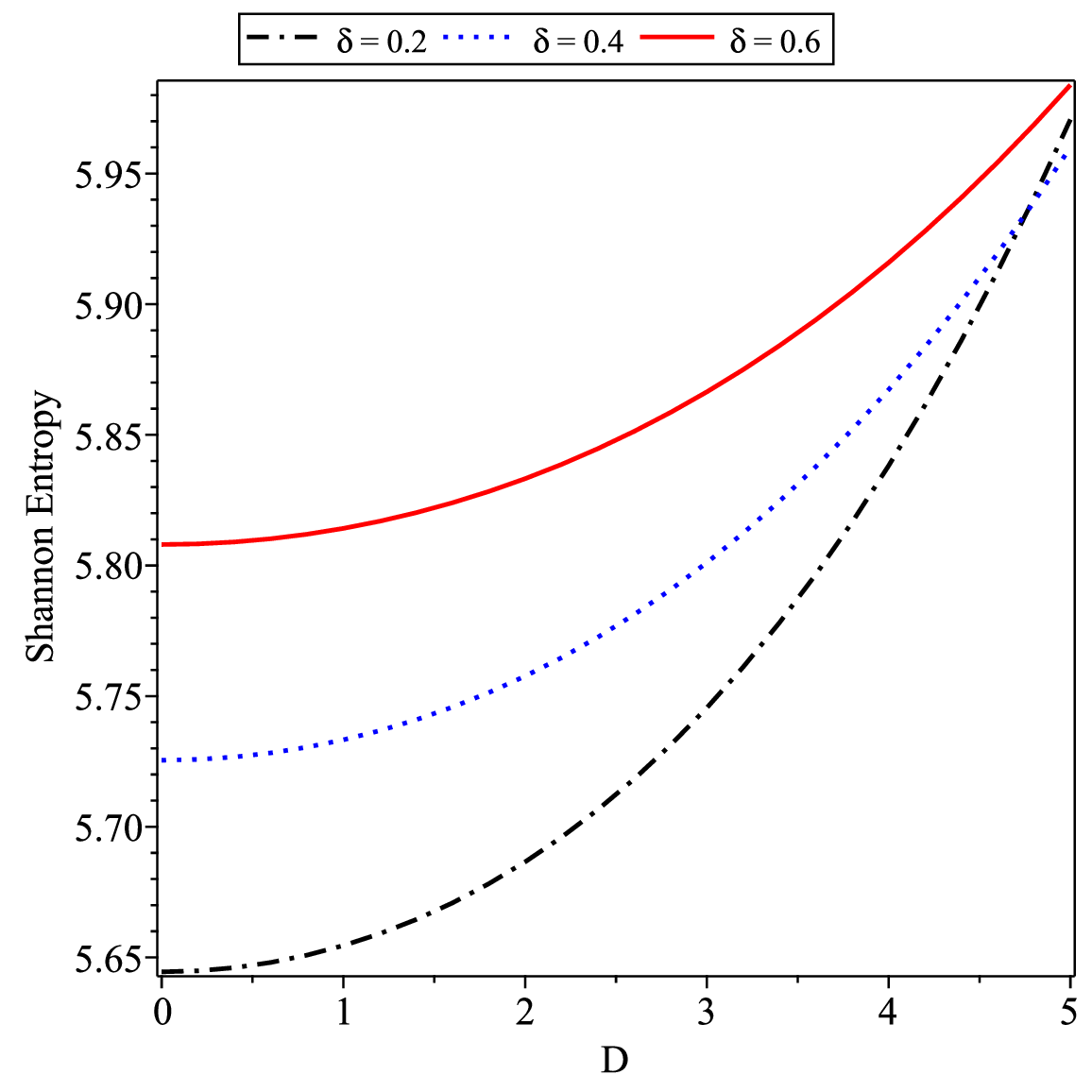}
  \caption{Shannon entropy versus the Dipole moment for $De=3$, $r_e=1$ (All quantities are in atomic units), with $n=2$ and $m=2$.}
  \label{fig:4}
\end{figure}
\begin{table}[htbp]
    \centering
   \captionsetup{format=plain,justification=centering}
    \caption{Fisher information and Shannon entropy for some diatomic molecules with $\delta=0.2, D=0.4$ in atomic units} 
    \label{tab:1}
\begin{tabular}{|c c c c c c c c|}
\multicolumn{2}{l}{} & \multicolumn{3}{l}{Fisher Information} & \multicolumn{3}{l}{Shannon Entropy} \\
\hline
           n        &  m  &  $I(Cs_2)$     &  $I(Li_2)$     &  $I(SiSn)$     &  $S(Cs_2)$     &  $S(Li_2)$     &   $S(SiSn)$    \\
\hline
\multirow{3}{*}{1}  & 0  &  1.20     &   3.01    &   6.32    &   6.2815    &  5.3274     &   4.6874    \\
                   &  1 &   1.15   &     2.86  &   6.46   &    6.3875   &   5.4561    &     4.7559  \\
                   &  2 &    1.13   &    2.75   &    6.52   &    6.5810   &   5.6816    &   4.8837    \\
\multirow{3}{*}{2}  & 0  &   1.39    &   3.34    &   8.20    &  6.9706     & 6.0591     &  5.2857     \\
                   & 1  &   1.32    &   3.14    &     7.80  &   7.0611  &   6.1641    &     5.3459  \\
                   &  2 &    1.25   &    2.91   &     7.73  &    7.2296   &    6.3581   &     5.4591  \\
\multirow{3}{*}{4}  & 0  &   1.25    &  2.84    &  8.11     & 8.0276     &   7.1660    &    6.2181  \\
                   & 1  &     1.18  &    2.74   &    7.84  &   8.0971   &    7.2455   &    6.2676   \\
                   &  2&    1.10   &    2.52   &    7.61   &   8.2311    &   7.3980   &     6.3601  \\
\multirow{3}{*}{6}  & 0  &   0.998    &  2.27    &  7.08    &   8.8416    & 8.0125     &   6.9510    \\
                   & 1  &    0.958   &    2.17   &  6.74   &    8.8981   &    8.0750   &     6.9925  \\
                   & 2  &    0.910   &    2.01   &    6.66   &   9.0096    &    8.2010   &    7.0710   \\
\multirow{3}{*}{8}  & 0  &   0.798    &   1.77   &   5.95    &  9.5087     &   8.7020    &    7.5595   \\
                   &  1 &     0.770  &   1.70    &   5.76    &   9.5562    &    8.7525   &     7.5960  \\
                   & 2  &   0.733   &    1.58   &   5.63    &    9.6512   &     8.8600  &     7.6640\\
\hline
\end{tabular}
\end{table}
\section{\label{sec:4}The Tsallis and The Renyi Entropies}
The Tsallis entropy  is a generalization of the Shannon entropy within the framework of non-extensive statistical mechanics. It is defined as follow:\cite{Kali2011}
\begin{equation}
T_q = \frac{1}{q - 1} \left(1 - W_q[\rho]\right),\quad q>0,\quad q\neq 1
\label{eq:39}
\end{equation}
The Renyi entropy serves as another extension of the Shannon entropy and is defined as:\cite{Kali2011}
\begin{equation}
R_q = \frac{1}{1 - q} \log \left(  W_q[\rho] \right), \quad q > 0, \quad q \neq 1
\label{eq:40}
\end{equation}
where
\begin{equation}
W_q[\rho]=\int \left[\rho(\vec{r})\right]^q \, d\vec{r}=\frac{N^{2q}}{2\beta^2}\frac{(2q-1)!!2\pi}{2^q q!}\int_{0}^{\infty} x^{q(2\lambda-1)+1}e^{-q x} \left[L_n^{2\lambda -1} (x)\right]^{2q}\, dx
\label{eq:41}
\end{equation}
In Eq.~(\ref{eq:41}) we have used the substitution $x=2\beta r$ and the fact that:
\begin{equation}
\int_{0}^{2\pi}\left[\Phi(\theta)\right]^{2q}\, d\theta=\frac{\left(2q-1\right)!!}{2^{q}q!}2\pi
\label{eq:42}
\end{equation}
To evaluate the integral on the right-hand side of Eq.~(\ref{eq:41}) it is useful to employ the linearization formula of Srivastava-Niukkanen for the products of various Laguerre polynomials:\cite{Sanchez2011}
\begin{equation}
x^\mu L_{m_1}^{(\alpha_1)}(t_1x)L_{m_2}^{(\alpha_2)}(t_2x)\cdots L_{m_r}^{(\alpha_r)}(t_rx)=\sum_{k=0}^{\infty} \gamma_k\left(\mu, \lambda, r, \{m_i\}, \{\alpha_i\}, \{t_i\}\right) L_k^{(\eta)}(x)
\label{eq:43}
\end{equation}
which is given in terms of the Lauricella’s hypergeometric functions of $(r + 1)$ variables as:
\begin{align}
& \gamma_k\left(\mu, \lambda, r, \{m_i\}, \{\alpha_i\}, \{t_i\}\right) \nonumber \\
& = (\lambda + 1)_\mu
\begin{pmatrix}
    \alpha_1 + m_1 \\
    m_1
\end{pmatrix}
\begin{pmatrix}
    \alpha_2 + m_2 \\
    m_2
\end{pmatrix}
\cdots 
\begin{pmatrix}
    \alpha_r + m_r \\
    m_r
\end{pmatrix} \nonumber \\
& \times F^{\left(r+1\right)}_A\left[\eta + \mu + 1, -m_1, \dots, -m_r, -k; \alpha_1 + 1, \dots, \alpha_r + 1, \eta + 1; t_1, \dots, t_r, 1\right]
\label{eq:44}
\end{align}
where the Pochhammer symbol $(a)_n=\frac{\Gamma(a+n)}{\Gamma(a)}$ and $\begin{pmatrix} \alpha_i + m_i \\ m_i \end{pmatrix}$ are the binomial coefficients. For the special case $(\eta = 0, \alpha_1 = \cdots = \alpha_r = 2\lambda-1, m_1 = \cdots = m_r = n, x = qt, t_1 = \cdots = t_r = \frac{1}{q}, \mu=q(2\lambda-1)+1, r = 2q)$, we obtain the linearization:
\begin{equation}
(qt)^{q(2\lambda-1)+1} \left[L_n^{(\alpha)}(t)\right]^{2q}=\sum_{k=0}^{\infty}\gamma_k\left(q(2\lambda-1)+1, 0, 2q, \{n\}, \{2\lambda-1\}, \left\{\frac{1}{q}\right\}
\right) L_n^{(0)}(qt)
\label{eq:45}
\end{equation}
The orthogonality property of the polynomial $L_n^{(\alpha)}(x)$ implies that the term with $k = 0$ is the only non-vanishing contribution to the integral Eq.~(\ref{eq:41}) \cite{Sanchez2011}. Thus, we have:
\begin{equation}
\int_{0}^{\infty} x^{q(2\lambda-1)+1}e^{-q x} \left[L_n^{\alpha} (x)\right]^{2q} \, dx=\frac{1}{q^{q(2\lambda-1)+2}}\gamma_0\left(q(2\lambda-1)+1, 0, 2q, \{n\}, \{2\lambda-1\}, \left\{\frac{1}{q}\right\}\right)
\label{eq:46}
\end{equation}
where:
\begin{align}
& \gamma_0\left(q(2\lambda-1)+1, 0, 2q, \{n\}, \{2\lambda-1\}, \left\{\frac{1}{q}\right\}\right)= \Gamma(q(2\lambda-1) + 2)
\begin{pmatrix}
   2\lambda + n -1 \\
    n
\end{pmatrix}^{2q} \nonumber \\
& \times F^{\left(2q+1\right)}_A\left[q(2\lambda-1) + 2, -n, \dots, -n, 0; 2\lambda, \dots, 2\lambda, 1; \frac{1}{q}, \dots, \frac{1}{q}, 1\right]
\label{47}
\end{align}
Hence:
\begin{equation}
W_q[\rho]=\frac{N^{2q}}{2\beta^2}\frac{(2q-1)!!2\pi}{2^q q!}\frac{1}{q^{q(2\lambda-1)+2}}\gamma_0\left(q(2\lambda-1)+1, 0, 2q, \{n\}, \{2\lambda-1\}, \left\{\frac{1}{q}\right\}\right)
\label{eq:48}
\end{equation}
so that the Tsallis and the Renyi entropies become:
\begin{equation}
T_q=\frac{1}{q-1}\left[1-\frac{N^{2q}}{4\beta^2}\frac{(2q-1)!!2\pi}{2^q q!q^{q(2\lambda-1)+2}}\gamma_0\left(q(2\lambda-1)+1, 0, 2q, \{n\}, \{2\lambda-1\}, \left\{\frac{1}{q}\right\}\right)\right]
\label{eq:49}
\end{equation}
\begin{equation}
R_q=\frac{1}{1-q}\log\left(\frac{N^{2q}}{4\beta^2}\frac{(2q-1)!!2\pi}{2^q q!q^{q(2\lambda-1)+2}}\gamma_0\left(q(2\lambda-1)+1, 0, 2q, \{n\}, \{2\lambda-1\}, \left\{\frac{1}{q}\right\}\right)\right)
\label{eq:50}
\end{equation}
\begin{figure}
  \centering
  \includegraphics[width=0.6\textwidth]{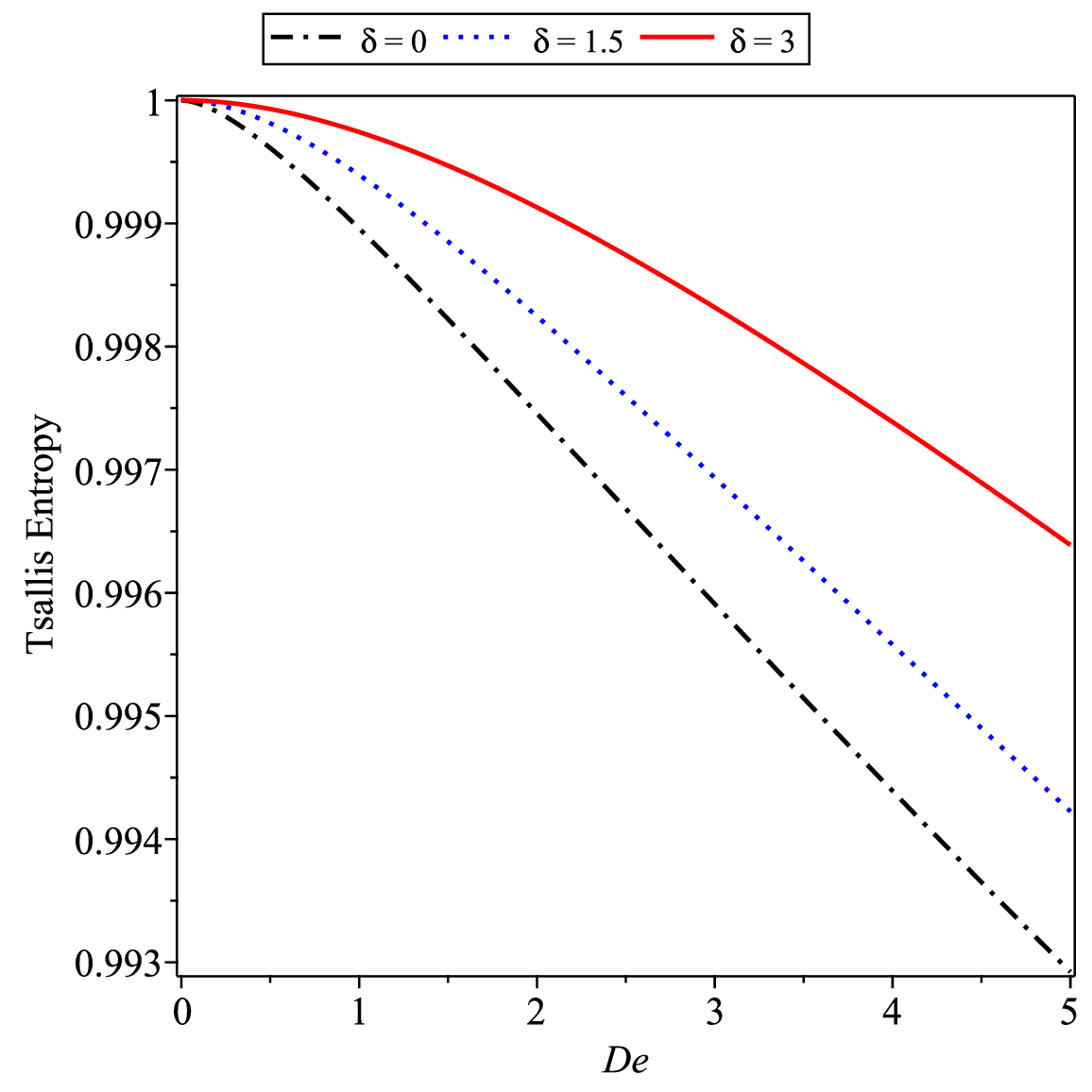}
  \caption{Tsallis entropy versus the Dissociation energy for $D=0$, $r_e=1$ (All quantities are in atomic units), with $n=2$ and $m=0$.}
  \label{fig:5}
\end{figure}
\begin{figure}
  \centering
  \includegraphics[width=0.6\textwidth]{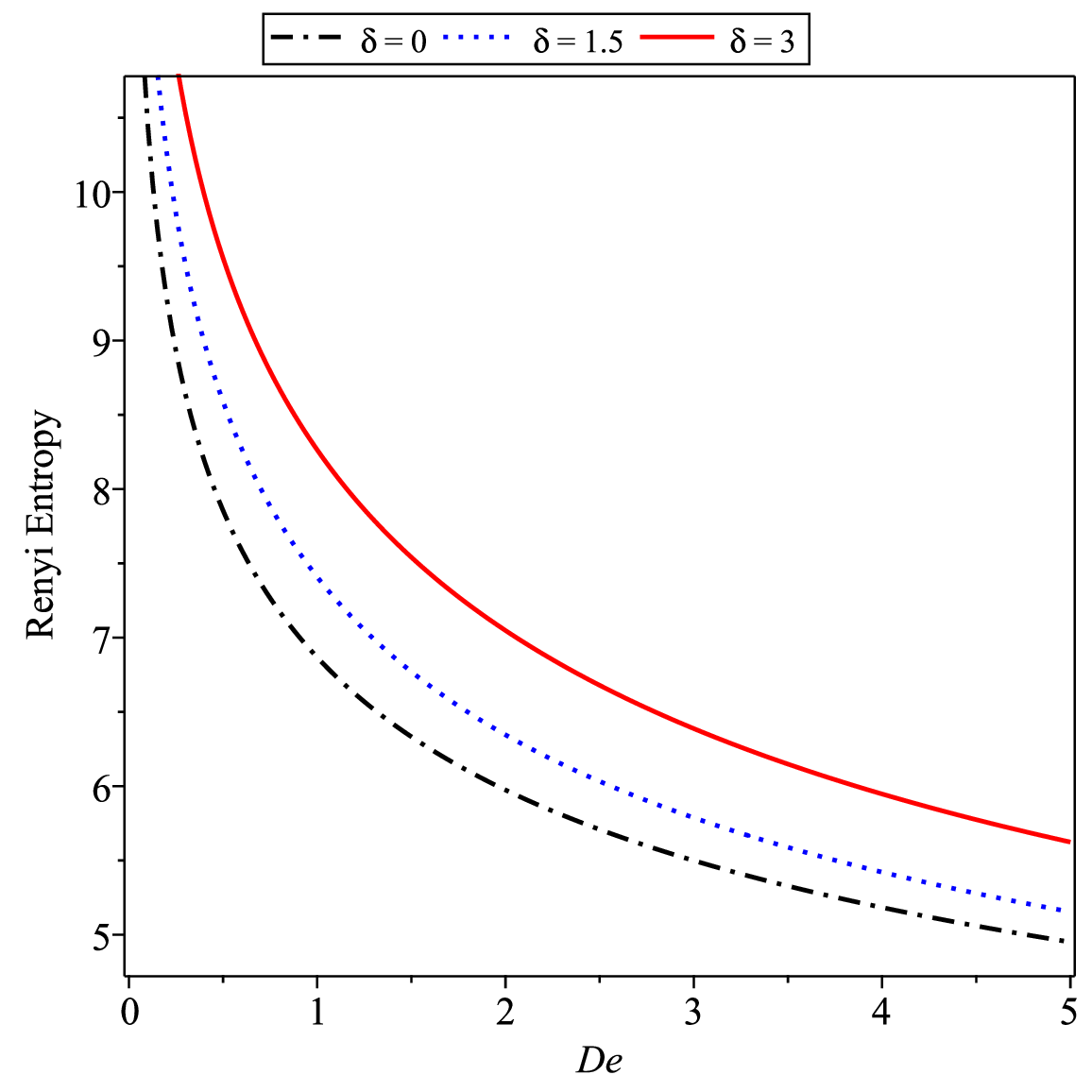}
  \caption{Renyi entropy versus the Dissociation energy for $D=0$, $r_e=1$ (All quantities are in atomic units), with $n=2$ and $m=0$.}
  \label{fig:6}
\end{figure}
\begin{figure}
  \centering
  \includegraphics[width=0.6\textwidth]{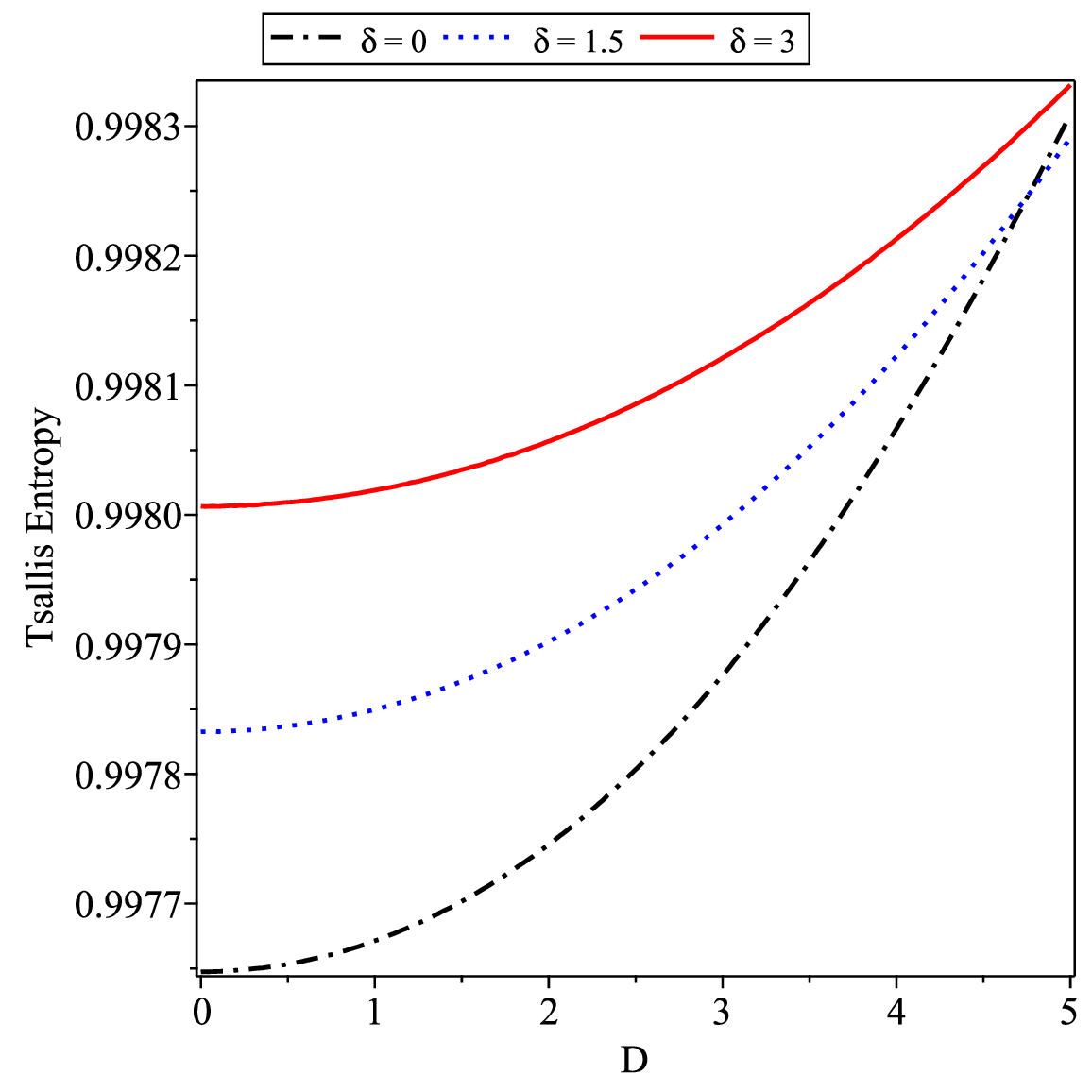}
  \caption{Tsallis entropy versus the Dipole moment for $De=3$, $r_e=1$ (All quantities are in atomic units), with $n=2$ and $m=2$.}
  \label{fig:7}
\end{figure}
\begin{figure}
  \centering
  \includegraphics[width=0.6\textwidth]{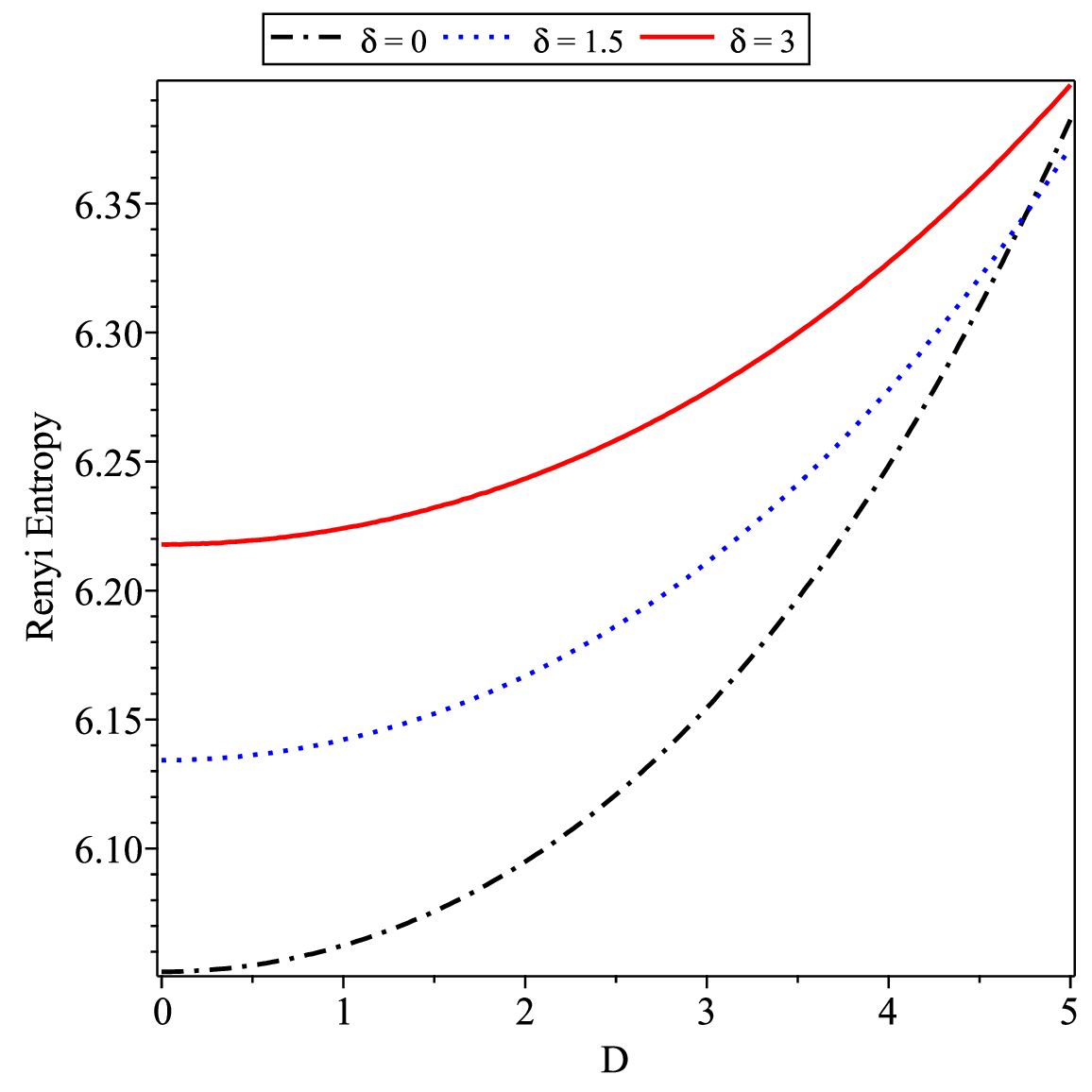}
  \caption{Renyi entropy versus the Dipole moment for $De=3$, $r_e=1$ (All quantities are in atomic units), with $n=2$ and $m=2$.}
  \label{fig:8}
\end{figure}
\begin{table}[htbp]
    \centering
   \captionsetup{format=plain,justification=centering}
    \caption{Tsallis and The Renyi Entropies for some diatomic molecules with $\delta=0.2, D=0.4$ in atomic units} 
    \label{tab:2}
\begin{tabular}{|c c c c c c c c|}
\multicolumn{2}{l}{} & \multicolumn{3}{l}{Tsallis Entropy} & \multicolumn{3}{l}{Renyi Entropy} \\
\hline
           n        &  m  &  $T(Cs_2)$     &  $T(Li_2)$     &  $T(SiSn)$     &  $R(Cs_2)$     &  $R(Li_2)$     &   $R(SiSn)$    \\
\hline
\multirow{3}{*}{1}  & 0  &  0.99751     &   0.99350    &   0.98784    &   5.9954    &  5.0362     &   4.4098    \\
                   &  1 &  0.99777    &     0.99430  &    0.98866   &    6.1056   &   5.1670    &    4.4792  \\
                   &  2 &   0.99816   &    0.99547   &    0.99003   &    6.2992   &   5.3962    &   4.6084    \\
\multirow{3}{*}{2}  & 0  &   0.99876    &   0.996882    &   0.99336    & 6.6897     &  5.7715     & 5.0144     \\
                   & 1  &   0.99887    &   0.99722    &     0.99375  &    6.7849   &   5.8848    &     5.0756  \\
                   &  2 &   0.99904   &    0.99771   &    0.99443  &    6.9526   &    6.0812   &     5.1898  \\
\multirow{3}{*}{4}  & 0  &   0.99957    &   0.99898    &  0.99740     &  7.7518     &   6.8894    &    5.9541   \\
                   & 1  &    0.99960  &    0.99907   &     0.99753  &   7.8273    &   6.9778   &    6.0038  \\
                   & 2 &    0.99965   &    0.99920   &    0.99775   &   7.9600    &   7.1311   &     6.0969  \\
\multirow{3}{*}{6}  & 0  &   0.99981    &   0.99957    &  0.99876    &   8.5686    &  7.7410     &  6.6907    \\
                   & 1 &    0.99982   &    0.99960   &    0.99881   &    8.6316   &    7.8141   &     6.7326  \\
                   & 2  &   0.99984   &   0.99964   &   0.99890   &   8.7413   &    7.9400  &    6.8115  \\
\multirow{3}{*}{8}  & 0  &   0.99990    &   0.99978   &   0.99933    & 9.2367     &  8.4331    &   7.3023   \\
                   &  1 &     0.99991  &   0.99980    &   0.99935    &   9.2909    &    8.4956   &    7.3386  \\
                   & 2  &    0.99992  &    0.99982   &   0.99939    &    9.3847   &     8.6028  &     7.4069\\
\hline
\end{tabular}
\end{table}
\section{\label{sec:5}Results and Discussion}
This section is devoted to the discussion of our numerical and graphical results. For convenience, we use the Hartree atomic units defined by $ \hbar = e = m_e = 4\pi\epsilon_0 = 1 $ (all figures in this article are drawn using atomic units). In Fig.~(\ref{fig:1}) we have examined the effect of dissociation energy $D_e$ on Fisher information for different values of the AB field parameter $\delta$. First, we mention that Fisher information is related to quantum kinetic energy, wherein the kinetic energy is decomposed into classical kinetic energy and purely quantum kinetic energy, the latter term being the Weizsäcker term, which is essentially identical to Fisher information \cite{Hamilton2010}. Fig.~(\ref{fig:1}) shows that the Fisher information increases with increasing dissociation energy $D_e$, implying a higher degree of localization. This is obvious since the dissociation energy is the amount of energy required to break the bond and separate the combining atoms in the molecule. In the framework of electronic structure, the dissociation energy is connected with the degree of electron localization within the molecule. A higher dissociation energy implies that the bond between the atoms within the molecule is stronger, resulting in electrons being more tightly held between the atoms, leading to a higher degree of localization. Moreover, Fig.~(\ref{fig:1}) shows that the Fisher information decreases with increasing AB field parameter $\delta$, indicating a decrease in localization.

In Fig.~(\ref{fig:2}) we have plotted the Fisher information against the dipole moment terms $D$ for different values of the AB field parameter $\delta$. In this figure, and elsewhere when we discuss the impact of the dipole moment on different information measures, we have to consider the accuracy of the characteristic number of the Mathieu function in Eq.~(\ref{eq:10}). Therefore, we have carefully selected the values for the parameters involved to ensure that the characteristic number remains consistent with the different leading terms. For $m=0, 1$, the accuracy of the characteristic number ~(\ref{eq:10}) is held for $b<1$.  For $m=2$, the accuracy can be extended to values of $b$ up to 20. The higher the angular number $m$, the wider the range of $b$  required for the accuracy of the characteristic number; In our case, we choose $m=2$. Fig.~(\ref{fig:2}) illustrates that increasing the dipole moment and the AB field parameter $\delta$ leads to a decrease in Fisher information, which results in a loss of information regarding localization.

In Fig.~(\ref{fig:3}) and Fig.~(\ref{fig:4}), we have depicted the influence of dissociation energy and dipole moment on Shannon entropy, respectively. The Shannon entropy gives insight into the precision and spatial localization of particles and it reflects the probability distribution and the stability of the system. A lower Shannon entropy indicates greater accuracy in predicting particle localization and, consequently, higher stability. Fig.~(\ref{fig:3}) illustrates that dissociation energy reduces the Shannon entropy, thus improving precision and spatial localization. This observation aligns with our previous findings, as the dissociation energy governs the bond strength between atoms in a molecule. Higher dissociation energy implies tighter electron binding between atoms, leading to increased localization. Fig.~(\ref{fig:4}) shows that the dipole moment increases the Shannon entropy and therefore decreases the precision and localization as expected. Moreover, for the Fisher information scenario, the AB field parameter $\delta$ negatively affects the accuracy of particle localization prediction. Shannon entropy increases with $\delta$, resulting in a loss of precision and increased uncertainty regarding particle localization in space, as shown in Figs.~(\ref{fig:3}) and ~(\ref{fig:4}).

We should note that we have used a numerical method to evaluate and plot the Shannon entropy in Figs.~(\ref{fig:3}) and ~(\ref{fig:4}). The reason for this is that the last integral of the Shannon entropy $S_4$, Eq.~(\ref{eq:38}), suits high radial numbers since it is calculated in the asymptotic limits \cite{Dehesa1998}.

In Figs.~(\ref{fig:5}) and ~(\ref{fig:6}), as well as Figs.~(\ref{fig:7}) and ~(\ref{fig:8}), we investigated the influence of dissociation energy and dipole moment on the entropies of Tsallis and Renyi, respectively. These entropies are just generalizations of the Shannon entropy. Figs.~(\ref{fig:5}) and ~(\ref{fig:7}) show that the decrease in the dissociation energy results in lower Tsallis and Renyi entropies, resulting in better precision and spatial localization. In contrast, increasing the dipole moment leads to higher Tsallis and Renyi entropies, thus reducing precision and spatial localization, as depicted in Figs.~(\ref{fig:6}) and ~(\ref{fig:8}). Additionally, an increase in the AB field parameter $\delta$ leads to a loss in precision and an increase in uncertainty regarding particle localization in space.

Tab.~(\ref{tab:1}) and ~(\ref{tab:2}) demonstrate the numerical analysis of Fisher information and Shannon entropy, and also Tsallis and Renyi entropies, for different eigenstates of some diatomic molecules, namely, $SiSn (X^3 Sigma_g^+: De=2.642965641 eV. re:=2.514 A^0)$, $ Li2 (X1 Sigma_g^+ : De=1.055918901 eV, re=2.6729 A^0)$ and $ Cs2 (X1 Sigma_g^+: De=0.4524686595 eV, re=4.648 A^0)$ \cite{Ciccioli2007, Kaur1998}. Shannon, Tsallis, and Renyi entropies are shown to increase with increasing radial number $n_r$ and angular number $m$, which means higher accuracy in predicting localization and stability. Fisher information decreases with increasing quantum angular number $m$, but when we look at the quantum radial number $n_r$, it increases and then decreases. The radial number that gives the maximum value to the Fisher information can be rounded analytically or numerically from Eq.~(\ref{eq:30}).
\section{\label{sec:6}Conclusion}
In this paper, we first derived the analytical solutions of the Schrödinger equation for a 2D system subjected to a non-central scalar potential and a vector potential of the Aharonov-Bohm (AB) effect and obtained the corresponding wave functions and energy eigenvalues. The non-central potential is composed of a Kratzer potential plus a dipole moment term. Subsequently, we conducted both analytical and numerical investigations into the information-theoretic measures, Fisher information, Shannon entropy, Renyi entropy, and Tsallis entropy.

The analysis of these information-theoretic measures was carried out for various parameters, including the dissociation energy $De$, the dipole moment $D$, the AB field parameter $\delta$, as well as the radial and angular quantum numbers for selected diatomic molecules. Our findings showed that Fisher information increases with an increase in dissociation energy $De$, while it decreases with increasing dipole moment $D$, AB field parameter $\delta$, and radial and angular quantum numbers.

In contrast, the Shannon, Renyi, and Tsallis entropies exhibit a decrease with rising dissociation energy $De$, and an increase with decreasing dipole moment $D$, AB field parameter $\delta$, and radial and angular quantum numbers. These observations collectively suggest that a higher dissociation energy enhances precision and particle localization in space. Furthermore, our observations indicate that increasing the dipole moment, AB field parameter, and radial and angular quantum numbers reduces the degree of particle localization.
\section*{Compliance with Ethical Standards}
\begin{itemize}
\item Funding: This work is supported by PRFU Research Project B00L02UN050120190001, Univ. Batna1, Algeria
\item Conflict of interest: The authors declare that they have no conflict of interest.
\item Ethical conduct: This study was carried out following the ethical standards of the journal.
\end{itemize}
\nocite{*}
\bibliography{InfforKratzer2DSystem}
\end{document}